\documentclass[a4paper]{jpconf}
\usepackage{graphicx,wrapfig}

\newcommand{\mgcs}{\ensuremath{\,\mathrm{mg/cm^{2}}}}

\newcommand{\micro}{\ensuremath{\mu}}
\newcommand{\alp}{\ensuremath{\alpha}}

\newcommand{\gam}{\ensuremath{\gamma}}

\newcommand{\kev}{\ensuremath{\,\mathrm{keV}}}
\newcommand{\mev}{\ensuremath{\,\mathrm{MeV}}}

\newcommand{\cm}{\ensuremath{\,\mathrm{cm}}}
\newcommand{\mm}{\ensuremath{\,\mathrm{mm}}}

\newcommand{\kv}{\ensuremath{\,\mathrm{kV}}}

\newcommand{\degr}{\hspace{0.2mm}\ensuremath{^{\circ}}}

\newcommand{\leff}{\ensuremath{\mathcal{L}_{\rm eff}}}		
\newcommand{\leffa}{\ensuremath{\mathcal{L}_{\rm eff}^{\alpha}}}	
\newcommand{\Er}{\ensuremath{{E}_{\rm r}}}		
\newcommand{\Eee}{\ensuremath{{E}_{\rm ee}}}		
\newcommand{\kevee}{\ensuremath{\,{\rm keV}\hspace{-0.4mm}_{\rm ee}}}
\newcommand{\mevee}{\ensuremath{\,{\rm MeV}\hspace{-0.4mm}_{\rm ee}}}
\newcommand{\kevr}{\ensuremath{\,{\rm keV}\hspace{-0.4mm}_{\rm r}}}
\newcommand{\Ynr}{\ensuremath{{Y}_{\rm nr}}}		
\newcommand{\Yer}{\ensuremath{{Y}_{\rm er}}}		
\newcommand{\Se}{\ensuremath{{S}_{\rm e}}}		
\newcommand{\Sn}{\ensuremath{{S}_{\rm n}}}		
\newcommand{\fp}{\ensuremath{{f}_{\rm p}}}		
\newcommand{\tauo}{\ensuremath{\tau_{\rm 1}}}	
\newcommand{\taut}{\ensuremath{\tau_{\rm 2}}}	
\newcommand{\A}{\ensuremath{A}}				
\newcommand{\B}{\ensuremath{B}}				
\newcommand{\CR}{\ensuremath{C\hspace{-0.4mm}R}} 
\newcommand{\iph}{\ensuremath{I\hspace{-0.4mm}P\hspace{-0.4mm}H}} 
\newcommand{\LET}{\ensuremath{L\hspace{-0.4mm}E\hspace{-0.4mm}T}} 
\newcommand{\dedx}{\ensuremath{{\rm d}E\hspace{-0.4mm}/\hspace{-0.4mm}{\rm d}x}} 

\begin{document}

\title{Study of nuclear recoils in liquid argon with monoenergetic neutrons}

\author{C.~Regenfus, Y.~Allkofer, C.~Amsler, W.~Creus, A.~Ferella, J.~Rochet, M.~Walter}

\address{Physik-Institut der Universit¬\"at Z¬\"urich, CH--8057 Z¬\"urich, Switzerland}

\ead{regenfus@cern.ch}

\vspace{-2mm}
\begin{abstract} 
For the development of liquid argon dark matter detectors we assembled a setup in the laboratory to scatter neutrons on a small liquid argon target. The neutrons are produced mono-energetically (E$_{\rm kin}$=2.45\,MeV) by nuclear fusion in a deuterium plasma and are collimated onto a 3'' liquid argon cell operating in single-phase mode (zero electric field). Organic liquid scintillators are used to tag scattered neutrons and to provide a time-of-flight measurement. The setup is designed to study light pulse shapes and scintillation yields from nuclear and electronic recoils as well as from \alp-particles at working points relevant to dark matter searches. Liquid argon offers the possibility to scrutinise scintillation yields in noble liquids with respect to the populations of the two fundamental excimer states. Here we present experimental methods and first results from recent data towards such studies.
\end{abstract}

\vspace{-9mm}
\section{Introduction}
Over the last years liquid xenon TPCs\,\cite{Xe100res, Xe100} established themselves among the leading techno\-logies for WIMP searches. Considerable effort is being made by various groups to bring the liquid argon (LAr) sector to a competitive level\,\cite{CRpatras2010}.  Liquid argon has the potential to be a large and sensitive multi purpose detector due to its low ionisation potential and large abundance on earth. Interest in the low energy frontier of this technology increased by recent progress in the production of $^{39}$Ar depleted argon from dwell gases. Here we describe research activities which are part of the design study for a next generation dark matter facility, DARWIN\,\cite{Darwin}, presently proposed as a combined liquid\,xenon\,-\,liquid\,argon installation. 

WIMPs, hypothesised to be distributed as a thermalised halo in our galaxy, should produce nuclear recoils in any target on earth, which can be detected and isolated in noble liquids through their characteristic excitation and ionisation patterns. However the complex microscopic processes which lead eventually to the scintillation and charge signals are not very well understood at low energies 
and are currently the subject of experimental controversy. The experimental determination of light and charge yields at keV energies and possibly below is therefore of highest interest for this research since it defines the energy scale and sensitivity of the experiment. The uncertainty in the signal calibration of nuclear recoils still yields the largest contribution to the systematical error for liquid xenon WIMP searches\,\cite{Xe100res,Plante} despite the large efforts of various groups\,\cite{Plante, Aaron} to determine this quantity. In the liquid argon sector the situation is even worse, only scarce information about the scintillation efficiency can be found in literature\,\cite{Warp,Mckin} and no comprehensive measurement of the charge yield exists at all. 

The energy dependent light yield \Ynr\ of nuclear recoils is commonly described by the unitless quantity \leff, the relative scintillation efficiency comparing to the light yield of recoiling electrons (\Yer) of same kinetic energies. By convention \leff\ is measured at zero electric field. The value of \leff\ is determined by the product of two fundamental quenching processes. Firstly towards low energies an increasing fraction of the recoil energy is lost to heat (nuclear quenching) as described by Lindhard\,\cite{Lind1,Lind2}. A second mechanism, less well understood, originates in interactions among the excited and ionised states created in the process of electronic stopping, depending on the density of production of these states (luminescence quenching). The light yield of recoiling electrons is not affected by nuclear effects and shows a linear response for energies above some tens of keV\,\cite{Mei} due to relatively low values for the ionisation density in this regime. This allows for the convenient comparison of the energy scales of both recoil types, where \Eee\ represents the (equivalent) energy of an electron to produce the same amount of light as a recoiling nucleus with energy \Er. Electron light yields are traditionally determined at the 122\kev\ line of $^{57}$Co. Electric fields reduce the recombination of free charge carriers thus decreasing scintillation light\,\cite{Reclum}. In the estimation of light yields this effect is taken into account by two energy and field dependent correction functions \Sn\ and \Se, for nuclear and electronic recoils, respectively, which are unknown for liquid argon. In liquid xenon\,\cite{Aprile06,Manz} values for \Sn\ are close to unity while \Se\ drops to below 0.5 at fields $>$1\,kV/cm. This is due to screening effects at the large ionisation densities of nuclear recoils. Thus the relative scintillation efficiency can be calculated by  \leff = \Ynr /\Yer\,$\cdot$\Se /\Sn = \Eee /\Er\,$\cdot$\Se /\Sn. In this work the values for \Se\ and \Sn\ are set to unity.  

Liquid argon allows for a separation of the scintillation signal in fast and slow components due to the large difference in their lifetimes. This effect is exploited for background discrimination in LAr dark matter detectors\,\cite{Hitachi}. Here we apply pulse shape analysis to light yield studies in regions of both, low and high ionisation densities in a noble liquid. In the following we present a brief overview of the experimental method and show first results obtained from data recorded under neutron, \alp- and \gam-irradiation in liquid argon at zero electric field. 

\vspace{-2mm}
\section{Experimental Setup}
\label{expsetup}

We induce nuclear recoils in the liquid argon cell by fast neutrons (2.45\mev) from a deuterium-fusion-generator made by the company NSD-Fusion, Germany. It delivers up to 2$\cdot$$10^6$  isotropically emit\-ted  mono-energetic neutrons from the two-body reaction dd\,$\rightarrow$\,$^{3}$He\,+\,n (dd\,$\rightarrow$\,t\,+\,p is 
\begin{wrapfigure}{r}{0.6\textwidth}
\vspace{-6mm}
\centerline{\includegraphics[width=0.6\textwidth]{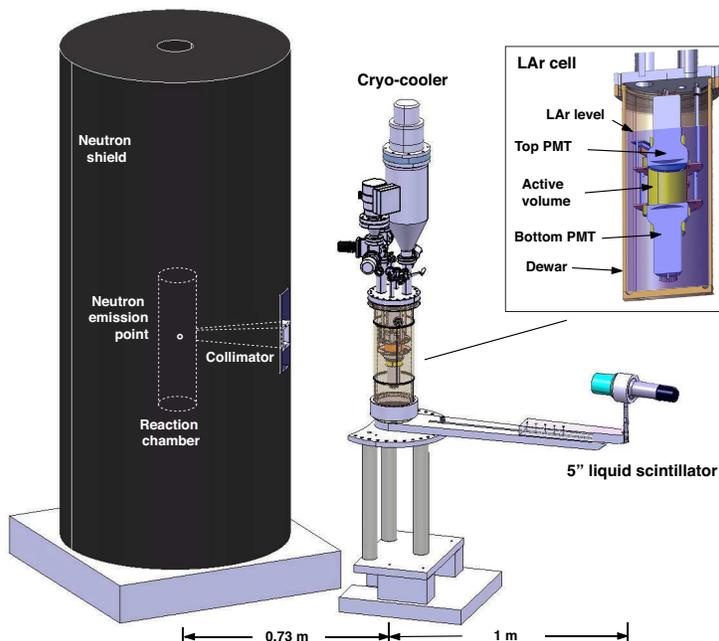}}
\vspace{-4mm}
\caption{Setup of the scattering experiment with neutron generator, shield and zoom on the LAr cell.}
\label{fig:setup}
\vspace{-4mm}
\end{wrapfigure}
equally probable). The overall setup was developed in collaboration with the producer. The fusion rate in the deuterium plasma is controlled by an electrical DC field generated by an adjustable constant current (1-15\,mA) HV supply. The voltage (30-120\kv) to keep the plasma current stable is regulated indirectly by the pressure of the deuterium gas which is released from heated getter disks storing deuterium on their surfaces. Higher voltage (lower gas pressure) produces higher fusion rates but also a harder bremsstrahlung background spectrum. The environment is shielded from neutron and X-ray radiation by a 1600\,kg poly\-ester cylinder with 2\mm\ Pb cladding (fig.\,\ref{fig:setup}) keeping the radiation dose far below the limit of 2.5$\mu$Sv/h imposed by CERN's radiation protection requirements. A safety area with access control and radiation interlock completes the setup. A polyethylene collimator with square cross section restricts the emission of neutrons in a solid angle of about 0.2\% of 4$\pi$, covering fully the sensitive volume of the (cryogenic) LAr cell. For the measurements described here the cell was located at a distance of about 73\cm\ from the neutron emission point while the liquid scintillator counter (LSC) could be set at various scattering angles on a 1\,m long arm rotatable around the centre of the cell to detect scattered neutrons. Errors on angles or mechanical misalignments were estimated to be below 0.5\degr . 

The internal structure of the cryogenic cell is shown in the inset of fig.\,\ref{fig:setup}. Two tetra-phenyl-butadiene (TPB) coated 3'' PMTs (Hamamatsu R6091-MOD) with bialkali photo cathodes and Pt underlay (QE$\approx$15\%) are arranged face to face at a distance of 47\,mm forming a cylindrical sensitive volume of roughly 0.2$\ell$. This volume is defined by a thin Al cylinder holding the TPB coated reflector foil\,\footnote{1\mgcs\ TPB on Tetratex; PMT coatings 0.08\mgcs\ TPB. See\,\cite{bocco,manu} for details.} to shift the VUV scintillation light to a longer wavelength. A small $^{210}$Po \alp-source of about 40\,Bq activity is installed in the centre of the cell. The source was coated with some tens of \micro m plastic (Paraloid B-72) to spread the energies of emitted \alp s over a broad range. The Al cylinder is polarised to the same voltage as the photo cathodes of the PMTs to keep the internal electrical field close to zero. A 60\,$\ell$/s turbopump is used to evacuate the chamber to typically $10^{-6}$\,mbar prior to filling with argon gas class 60 (impurities\,$\leq$\,1.3\,ppm). A membrane pump provides for continuous recirculation via two in parallel mounted OXISORB-W cartridges which reduce the O$_{2}$ and H$_{2}$O levels to $<$5 and $<$30\,ppb, respectively. The gas is condensed on top of the chamber at the cold head of a Sumitomo CH210 cryocooler system with about 80\,W cooling power driven by a Sumitomo F-70H helium compressor. 

A LabView based slow control system regulates the cold head temperature and records temperatures, pressures and liquid levels. The analogue signals from the two PMTs are each split into 2 inputs of a LeCroy WavePro\,735Zi DSO, sampled with 5000 points at 1\,GS/s and stored to the hard drive. The splitting of the signals permits a large dynamic range to fully reconstruct \alp-particles while maintaining high signal over noise ratios for single photons. Cosmic signals however can saturate the FADC\,\cite{CRpatras2010}. Coincidences between PMT signals and external detectors were achieved with the programable trigger logic in the oscilloscope. Signals of neutrons in the LSC were processed in a dedicated analogue pulse shape discriminator (Mesytec MPD4\,\cite{mesy}) and fed into the external input of the oscilloscope.

\vspace{-2mm}
\section{Data reconstruction and light yield}
\label{calib}

Scintillation signals in liquid argon are determined by numerical integration of the digitised photo currents of the two PMTs, normalised to their mean single photon charges. Due to the generally long integration times in LAr an iterative method was developed to calculate precise 
\begin{wrapfigure}{r}{0.60\textwidth} 
\vspace{-6mm}
{\includegraphics[height=4.4cm]{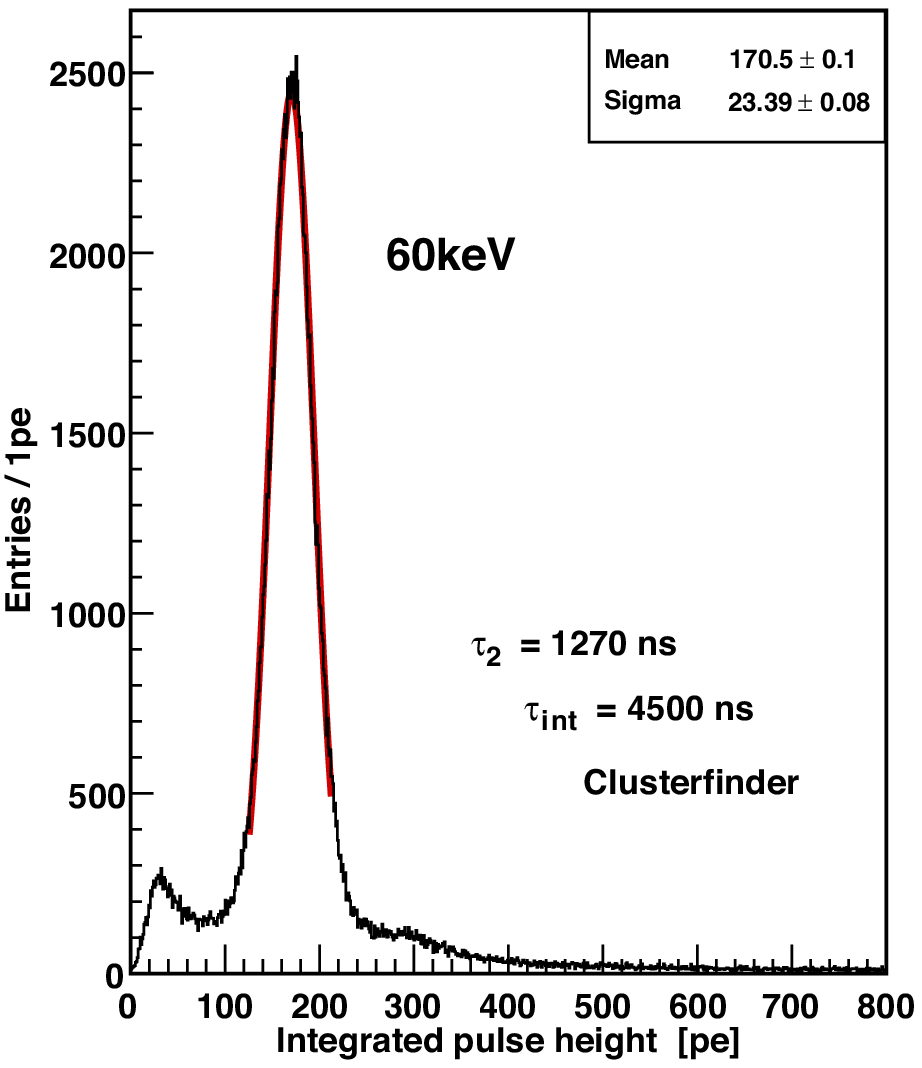}\hspace{0.5mm}
\includegraphics[height=4.4cm]{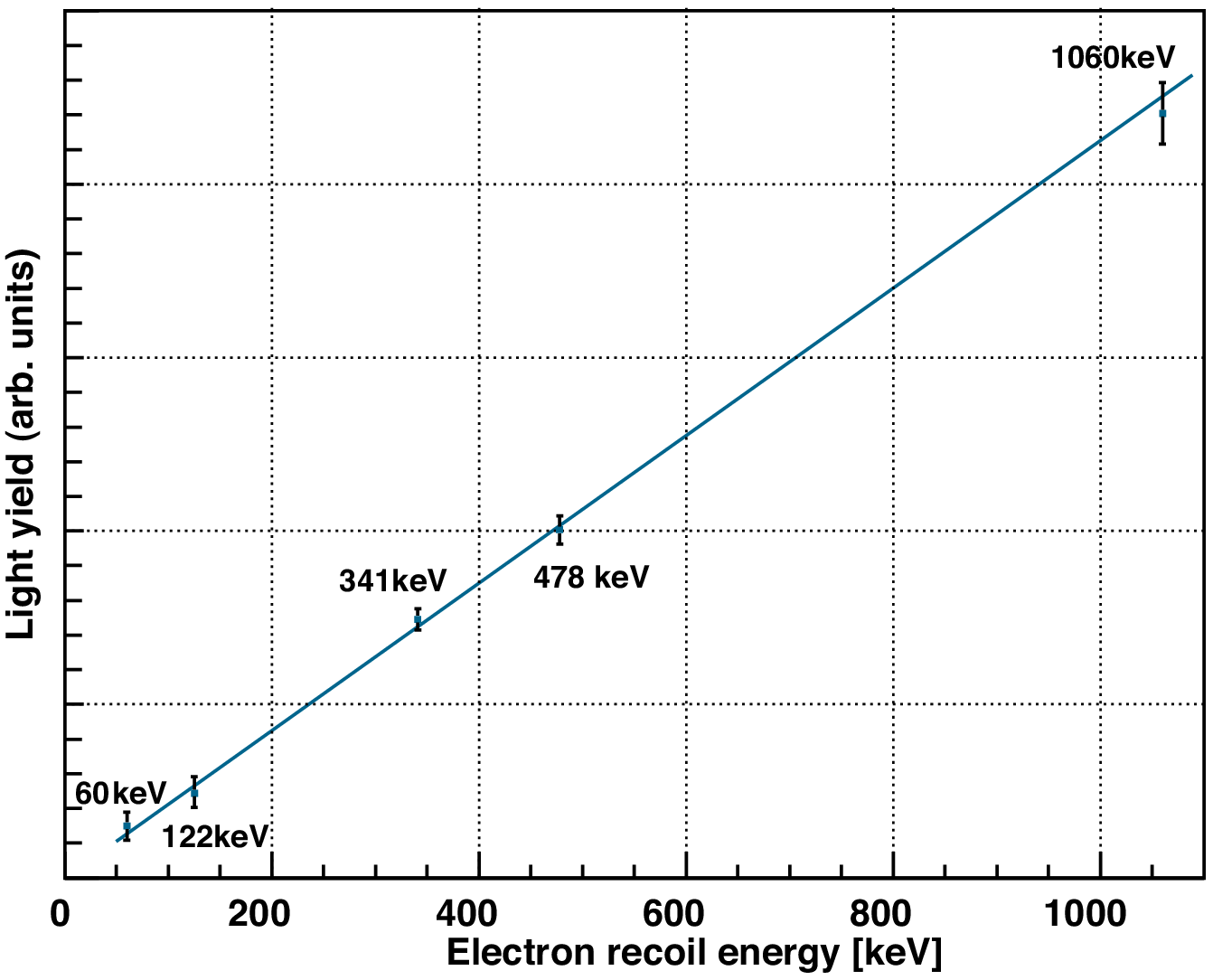}}
\vspace{-4mm}
 \caption{\small\sl Left: 60\,keV line from the raw signals of the $^{241}$Am data. 
 Right: light yield of various photon sources.}
\label{fig:ly}
\vspace{-5mm}
\end{wrapfigure}
pedestal values on an event-by-event base. The sum of both channels yields the raw integrated pulse height (\iph) in units of photo electrons (pe). This value is furthermore corrected for the finite integration time, losses by software thresholds (cluster finder) and most importantly by losses due to impurities in the liquid argon (see next section). Light yield calibrations were performed periodically during data taking by means of a strong external $^{241}$Am source producing a prominent 60\,keV photo peak in the integrated pulse height spectrum of the data (fig.\,\ref{fig:ly} left). 
The gains of the PMTs were determined from distributions of LED pulses or single photon signals collected from event tails in the data. The mean single photon charge is calculated adding 14.5\% to the most probable value of these distributions to take in account their skewness. The gains during several weeks of data taking in summer 2011 were found to be stable within 2.7\% leading to an average light yield of $3.75\pm0.1$\,pe/keV. To check the linearity of the system we employed various external photon sources, $^{57}$Co (122\,keV photo peak), $^{22}$Na (511 and 1275\,keV Compton edges), as well as $^{137}$Cs (662\,keV Compton edge). Figure\,\ref{fig:ly} (right) shows a linear fit to the raw measurements taken under similar conditions. 

\vspace{-3mm}
\section{Impurity effects --  light yield corrections}
\label{impquenching}

In liquid argon VUV fluorescence (128\,nm) from the so-called second continuum is the dominant mechanism  for light emission under excitation. The light pulse shape is well described by the sum of two exponentials originating in the radiative decays of two fundamental excimer states. Atomic selection rules are the cause for a large difference in their lifetimes, \tauo\ and \taut, approximately 6 and 1600\,ns, for singlet and triplet states, respectively. This feature is used for background discrimination in LAr dark matter detectors thanks to the ionisation density effect\,\cite{Hitachi} modifying the pulse shape according to specific energy losses of particles. Suppression factors up to 3 orders of magnitude are reached\,\cite{Lippi1} for recoil energies above 30\kevr\ using a likelihood based discrimination method. The improvement of such algorithms is also part of our research but will be described elsewhere. 

In the following we decompose individual or averaged signal traces by summing two exponential decays convoluted with a gaussian which describes general time spreads (2.9\,ns). We denote the integrals of fast and slow scintillation components from the singlet and triplet states with \A\ and \B, respectively, $Y$=\A+\B\ being the total mea\-sured light yield. Furthermore we denote the fraction \A /(\A +\B) as the component ratio \CR, representing the relative strength of the fast portion of the scintillation light. We want to stress that values for \CR\ are similar but somewhat different from the common variable for the prompt light fraction \fp, determined from short (typ.~50\,ns) and long signal integrations (typ.~4500\,ns).

Due to their long lifetime triplet states undergo various collisions with neighbouring particles before they eventually decay. Thereby interactions with impurities can cause the (non-radiative) destruction of these states and hence induce losses in the scintillation light (impurity 
\begin{wrapfigure}{r}{0.52\textwidth} 
\vspace{-5mm}
\centerline{\includegraphics[height=4.8cm]{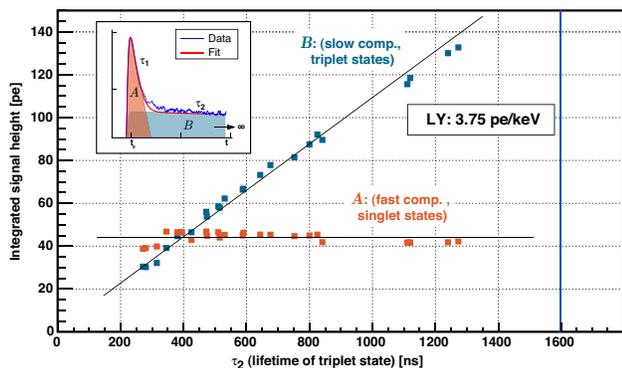}}
\vspace{-4mm}	
\caption{\small\sl Yield of the fast and slow scintillation components under different purity conditions.}
\label{fig:singtrip}
\vspace{0mm}
\end{wrapfigure}
quenching\,\cite{lumquench,nitoxy}). This process occurs in com\-petition to the natural decay of the states and results in an apparent reduction of the triplet lifetime \taut\ in the 
data\,\footnote{In LAr the value of \taut\ can be used to estimate purity levels in respect to scintillation yields\,\cite{ArdmFirst}. In our setup purity levels are predetermined by residual H$_{2}$O on internal surfaces of the vacuum and gas system.}. Taking into ac\-count the nor\-malisation factor for the exponential PDF the corrected light yield\,\cite{lumquench} is found from the relation
\vspace{-2mm}
\[ Y_{\rm cor}=\A\,+\,\B_{\rm cor}=\A\,+\,\B/\taut\,\,\cdot\,\tau^{\rm max}_2\,\, ,
\vspace{-2mm} \]
where $\B_{\rm cor}$ is the extrapolated yield of the triplet component and $\tau^{\rm max}_2$ the undisturbed lifetime of triplet states in LAr (1600\,ns\,\cite{Hitachi,ArdmFirst}). The fraction B/\taut, corresponding to the triplet decay rate at time zero, is effectively invariant vs.~variations of purity and completely determined by the fit. In the same way the value for the corrected component ratio is given by \CR$_{\rm cor}$\,=\,\A /(\A +\B$_{\rm cor}$). Corrected light yields are conveniently calculated by $Y_{\rm cor}$\,=\,\A/\CR$_{\rm cor}$. 
As an example fig.\,\ref{fig:singtrip} shows the two components of such fits to averaged signal traces of 60\,keV calibration data versus the value found for \taut\ under different purity condition. Values for \A\ remain essentially unaffected (above \taut\,$>$\,400\,ns) while \B\ increases proportionally the measured \taut. From these results we conclude that other quenching mechanisms, such as VUV absorption by impurities, are not affecting light yields under the present geometrical conditions and purity levels. All yields as well as \CR s presented in the following are corrected for light losses according to the method presented here.

\vspace{-4mm}
\section{Time structure of the scintillation light}
\label{cr}

The exponential decay model assumes production and modification of excimer states being completed on time scales shorter than the measured decay constants. Moreover the experi\-mentally observed  invariance of lifetimes, i.e.~their independence on ionisation densities, rules out\,\cite{Hitachi2} the participation of these states in the (fast) processes of luminescence quenching during production\,\footnote{In contrast, quenching by impurities is a relatively slow  process of destruction of (long lived) excimer states.}. However, in the intermediate region between fast and slowly decay modes, the two component fit systematically underestimates the data. The situation can be  improved by the introduction of a third component. Its relative contribution scales with the value for \CR\ and ranges between 3 and 7\% of the total integral\,\cite{phdvitto}, while lifetimes may vary between 50 and 200\,ns. The nature of this effect is presently not understood, possible explanations comprise PMT-after-glow and recombination light. In this work we exclude the transition region from the fit, pending a systematic study on this effect, e.g.~by measuring its electric field dependence.

In the following we compile measurements of \CR s determined from event-by-event likelihood fits for recoiling Ar nuclei, \alp-particles and electrons. The former were induced by 2.45\,MeV neutrons from the generator, the second were emitted from the built-in $^{210}$Po source, and the latter were induced by 511\,keV photons from an external $^{22}$Na source. As already mentioned the $^{210}$Po source was coated with a thin layer of plastic to degrade and spread the energy of emitted \alp s. To exclude bias in the reconstruction of \CR\ we tested the software with simulated pul\-ses of fixed \CR\ in a large variety of sizes. No significant deviation of the expected (flat) res\-ponse could be found. Data for neutrons and \alp s were triggered only on signals in the liquid argon, while an additional coincidence with an external detector was required for the photon data to tag the emission of two 511\,keV photons. This data was also used to determine the\,\,time calibration for the time-of-flight (TOF) measurement for scattered neutrons as well as the trigger roll-off at small pulse heights. The latter is derived from the lower end of the Compton  
\begin{figure}[h] 
\vspace{-2mm}
 \centering
 \includegraphics[height=6cm]{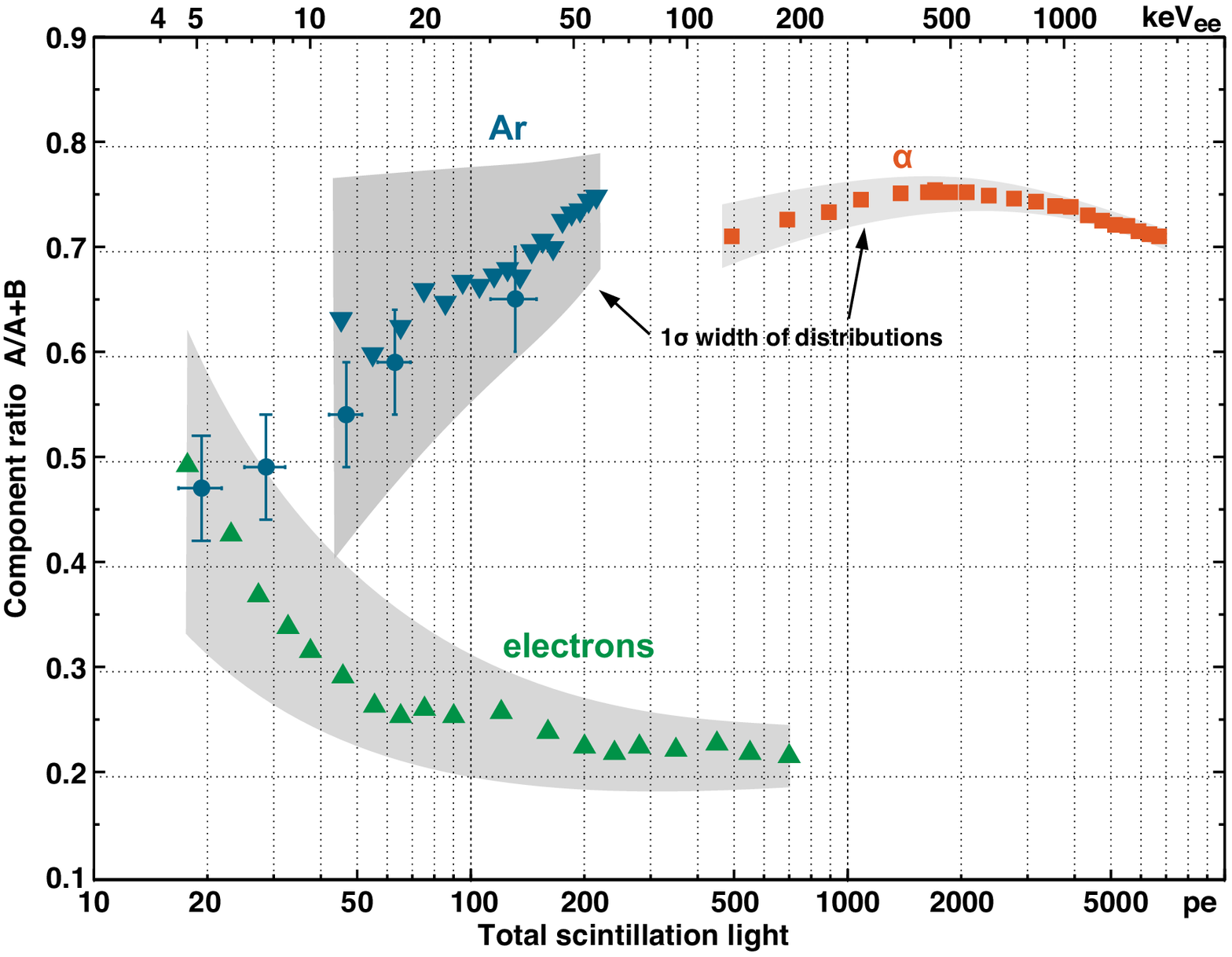}\hspace{6mm}
 \includegraphics[height=6cm]{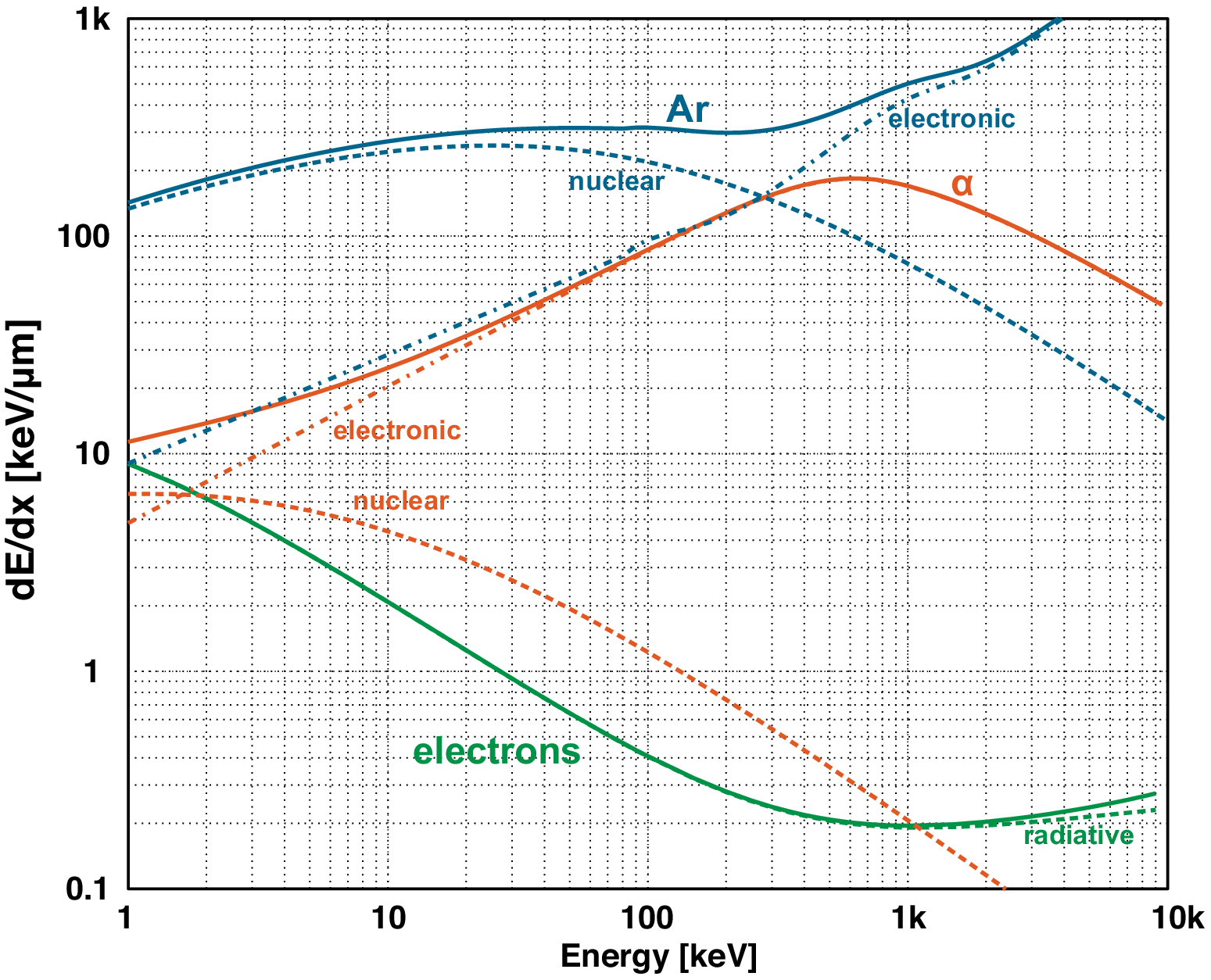}
 \vspace{-6mm}
 \caption{\small\sl Left: \CR\ vs.~light yield for Ar-, \alp- and e$^{-}$-recoils. The grey zones indicate the 1$\sigma$ spread. Right:~total stopping power (solid) and contributions (dashed) from SRIM, ASTAR and ESTAR\,\cite{Star}.}
    \label{fig:stopping}
    \vspace{-4mm}
\end{figure}
spectrum. While photon and \alp-data are practically background free, nuclear recoils suffer from a contamination with inelastic neutron collisions with argon nuclei producing photons. However due to the small geometric dimensions and relatively low kinetic energies of the neutrons these events are estimated not to contribute more than 7\% to the data. A Monte Carlo study is in progress.

Figure\,\ref{fig:stopping} (left) shows the values for \CR\ vs.~the total amount of scintillation light produced by these particles. The two prominent bands from electrons (upward triangles) and nuclear recoils (downward triangles) merge for energies below about 10\kevee. The grey zones show the 1$\sigma$ widths of the distributions, derived from gaussian fits of vertical slices. Errors are dominated by systematics introduced by the analysis method and are estimated to be roughly 0.25 of the distribution widths. The values at low energies for argon recoils are also shown (circles) and derived from fits to mean traces (see section\,\ref{leff}). The right plot shows the total stopping power for the same particles as well as the nuclear and electronic contributions to it. The data points were collected from the websites of SRIM, ASTAR and ESTAR\,\cite{Star}, respectively.

The correlation between both plots is obvious.  Larger ionisation densities generally lead to stronger interaction among participating particles producing a larger fraction of singlet states in the scintillation light. This is also true for nuclear recoils towards low energies as well as for the maxima in the curves for \alp-particles. Interesting is furthermore the value close to 0.25 for \CR\  
at the lowest ionisation densities for electrons. This can be interpreted as the isolated production of singlet and triplet excimer states (no interactions among each other) according to their statistical weights. In this case a factor of 1:3 is expected for the ratio in their populations. For the upper mentioned arguments of lifetime invariance we can exclude quenching effects by interactions among the excimer states after production. Therefore deviations from the minimal value of 0.25 either imply the enhanced production of singlet states on the exciton level or the transfer from triplet to singlet states by collisional spin change shortly after production in areas of high ionisation density. 

The physical relation between the population of excimer states at different specific energy losses can be unfolded from the measurement taking into account the integral over the whole trajectory. In the next section we determine the energy scale for \alp-particles by using a simple linear model for the relation between \CR\ and linear energy transfer. 

\vspace{-3mm}
\section{Scintillation yield of \alp-particles in LAr}
\label{leffa}

In this section we extract the relative scintillation yield \leffa\ of \alp-particles by comparing the shapes of the \CR\ to the linear energy transfer (\LET). In good approximation the latter is assumed to be identical with the electronic stopping power \dedx\,($E$).
For the fit function we 
\begin{wrapfigure}{r}{0.54\textwidth}
\vspace{-4mm}
\centerline{\includegraphics[height=5.9cm]{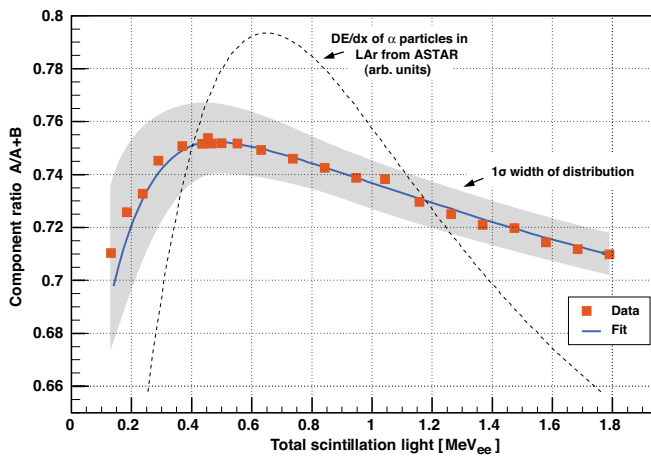}}
\vspace{-3mm}
 \caption{\small\sl Component ratio of \alp-particles in LAr and fit vs.~total 
 scintillation light in units of \mevee.}
 \label{fig:alphaquench}
\vspace{-1mm}
\end{wrapfigure}
use a 10$^{\rm th}$ degree polynomial parametrisation of the ASTAR curve from fig.\,\ref{fig:stopping}. The value for \CR\ is then calculated from the integral over a linear relation to the \dedx\ curve by
\vspace{-2mm}
\[ \CR(E)=\int_0^{x_{E=0}}\frac{{\rm d}\CR(E)}{{\rm d}E}\cdot\dedx\,(E)\,{\rm d}x ,\]
\vspace{-1mm}
\[{\rm with}\,\,\,\,\frac{{\rm d}\CR(E)}{{\rm d}E}= p_{1}\cdot\dedx\,(p_{2}\cdot E)+p_{3} ,\vspace{2mm}\]
\noindent where $E$ is energy, $x$ the path of the \alp-particle and $p_{i}$, $i$=1..3 are three free fit parameters. While $p_{3}$ describes an empirical vertical offset, $p_{2}$ is the parameter of interest describing the scale for the energy dependent yield (quenching). The simple comparison of the maxima in fig.\,\ref{fig:stopping} left and right gives a first estimate on the value \leffa\,$\cong$\,0.75\,. Figure\,\ref{fig:alphaquench} shows the measurements of \CR\ obtained by event-by-event likelihood fits of the signal shape from \alp-particles. The 1$\sigma$ spread is again illustrated by the grey band. The fit (solid line) describes the data rather well but starts to deviate at low energies probably due to our simple minded linear model. Also shown is the curve of the electronic stopping power for \alp-particles in LAr from ASTAR (dotted line), which actually deviates by less than 2\% from the curve for the total stopping in this energy range. 

A value of 0.74$\pm$0.04 is found for the parameter $p_{2}$ from the best fit. This corresponds to an estimate for the mean relative scintillation efficiency for \alp-particles in LAr in the range  0.18\,-\,2.5\,MeV. The result agrees within the errors with the value of 0.71 given in\,\cite{Mei} for 5\,MeV \alp-particles, assuming a nuclear quenching of 0.98. A similar value was also obtained in\,\cite{Hitachi2}. Due to their relatively small mass \alp-particles in this energy range are hardly affected by nuclear quenching. The light loss of roughly 25\% is hence entirely due to luminescence quenching.  

\vspace{-2mm}
\section{Light yield of nuclear recoils in LAr}
\label{leff} 

In this section we present the data taken in the experimental arrangement of the neutron generator described in section\,\ref{expsetup}\,. A 5'' liquid scintillator (LSC) was used for triggering, positioned under the angles of 30, 40, 50, 60 and 90\degr\ corresponding to 16.4, 28.5, 43.4, 60.5 and 120\,keV recoil energies, respectively. Due to the finite sizes of the detectors scattering angles are  
\begin{wrapfigure}{r}{0.59\textwidth}
\vspace{-5mm}
\centerline{\includegraphics[height=8.6cm]{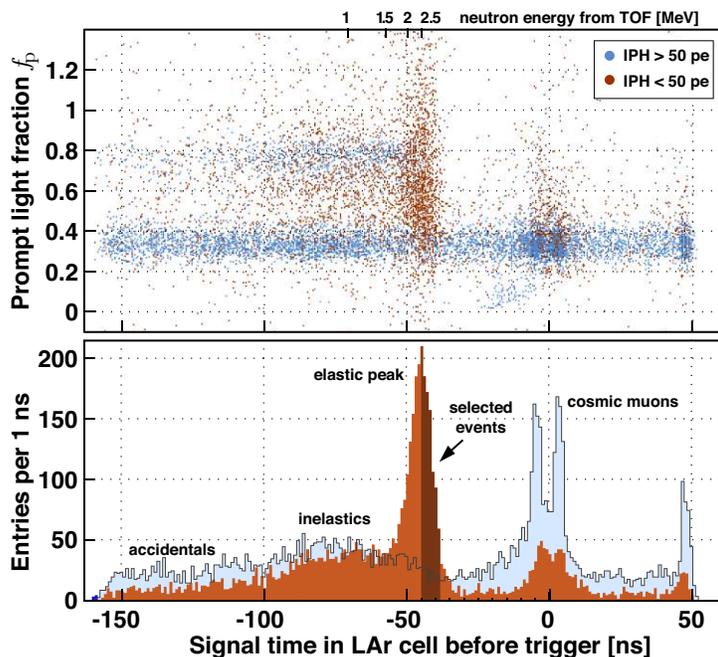}}
\vspace{-3mm}
 \caption{\small\sl \fp\ vs.~signal time in LAr with projection on the t-axis, for neutrons triggering the liquid scintillator counter.}
 \label{fig:30degdata}
\vspace{-5mm}
\end{wrapfigure}
distributed around the nominal angle roughly box-like (FWHM$\approx$10.8\degr). In most cases the neutron generator was operated at 80\,kV and 10\,mA, corresponding to the emission of about 2$\cdot$$10^5$ neutrons/s into 4$\pi$. The settings were chosen such that background induced by bremsstrahlung (accidental coincidences) is kept at an acceptable level. Under these settings about 1 neutron/min scatters off an argon nucleus in the active volume and is detected in the liquid scintillator 1\,m downstream. The direct line of sight between the exit of the collimator and the LSC was obstructed with a 20\,cm thick sheet of polyethylene. Background rates roughly amount to 5/min, originating mainly from two sources, cosmic muons saturating the LSC output and faking a neutron signal in the analog pulse shape discriminator MPD4, and accidental coincidences between diffusively scattered neutrons and bremsstrahlung. Events in the LAr cell are accepted when both PMTs show signals above 0.2\,pe in a 50\,ns time coincidence. A trigger is generated when an accepted event from the liquid argon cell occurs in a window of -150...+50\,ns around the arrival time of a confirmed neutron (MPD4) in the LSC. 

As an example the 30\degr\ scattering data (19k events) is shown in fig.\,\ref{fig:30degdata}. Events are printed in different colours according to their size, blue and brown for \iph$>$50\,pe and 
\iph$<$50\,pe respectively. The upper plot shows the prompt light fraction \fp\ vs.~the time difference (TOF) between the signals in the LAr cell and the LSC. Here \fp\ is calculated by the ratio of the scintillation light in the first 50\,ns to the total light. The histogram on the bottom shows the projection to the $t$-axis of the same events. The distribution of small signals (brown) is dominated by elastically scattered neutrons around the nominal flight time at  -45\,ns. Also visible is a broad distribution of inelastic scatters due to neutrons loosing a substantial fraction of their energy. A small uniform background originates in accidental coincidences of diffusely scattered neutrons with bremsstrahlung in the outer LAr volume. 

Two main cuts are applied to select events. Firstly a loose limit on \iph\ of 50\,pe (brown events) removes very efficiently accidental coincidences with photons, inelastic scatters with large energy deposit in the LAr cell, as well as cosmic muons. Secondly a time cut of a typically 5\,ns wide window to the right of the most probable value of the elastic scatter peak (dark area) serves 
\begin{wrapfigure}{r}{0.48\textwidth}
\vspace{-4mm}	
\centerline{\includegraphics[height=4cm]{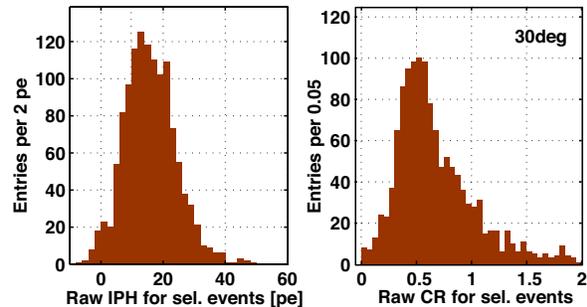}}
\vspace{-3mm}
 \caption{\small\sl Raw distributions of \iph\ and \CR\ after the time and 
 \iph\ cut for the 30\degr\ data.}
 \label{fig:30degsel}
\vspace{-3mm}
\end{wrapfigure}
as main event selector. We estimate roughly 15\% background  (incl.~mult.~scatters) in the selected data set originating mainly in inelastic scatters with small energy deposit in the LAr. Depending on the amount of data taken we end up with about 1k events per scattering angle shown in fig.\,\ref{fig:30degsel} for the 30\degr\ data. From the corresponding distributions we determine \A\ and \CR, the former from the distribution mean and the latter by a fit to the averaged signal trace of the selected events. A more precise study by comparing the measured distributions to the ones generated by Monte Carlo is in progress.  From a preliminary analysis of the 5 scattering angles we determine light yields as well as component ratios with errors of typically 15\%. 

\begin{figure}[h]
	\vspace{-4mm}
	\centering
	\includegraphics[height=5cm]{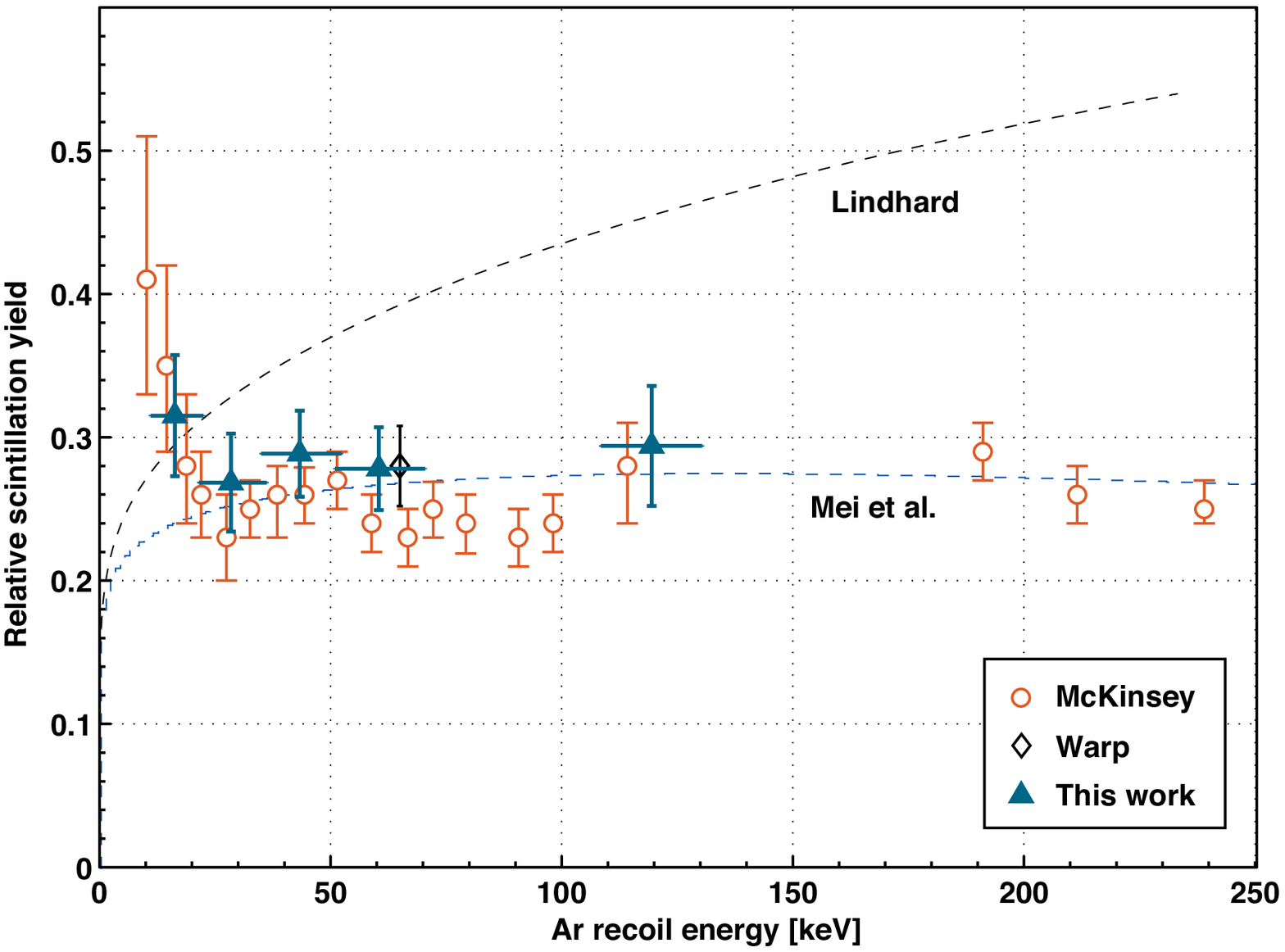}\hspace{12mm}
	\includegraphics[height=5.2cm]{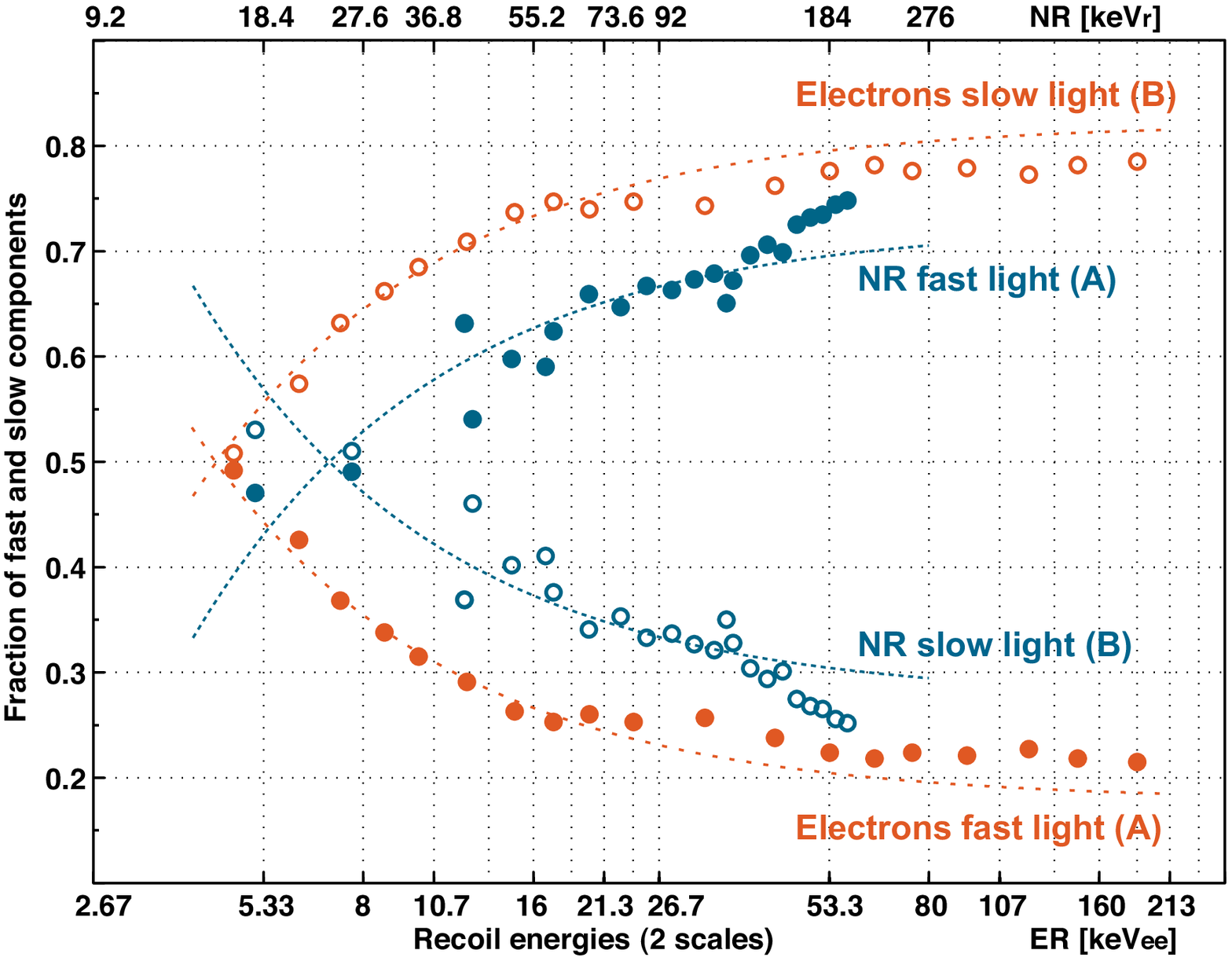}
	\vspace{-3mm}
	\caption{\small\sl Left: values for \leff\ with a comparison to theoretical curves and other published data. Right: relative scintillation components for nuclear recoils and electrons plotted against the two energy scales.}
	\label{fig:leff}
	\vspace{-4mm}
\end{figure}

Figure\,\ref{fig:leff} (left) shows our measurements together with the data from\,\cite{Mckin} and the one averaged value from\,\cite{Warp}. Our measurements are compatible with a flat interpretation of \leff, leading
to a mean value of $<$\leff$>$\,=\,0.29$\pm$0.03 for nuclear recoils at energies above 20\kevr\ or 6\kevee. With the present data and analysis we can neither exclude nor confirm the  increase of \leff\ at low energy from\,\cite{Mckin}. A comparison with the theoretical description of luminescence quenching in LAr by a simple saturation law combined with the Lindhard model (Mei et al.\,\cite{Mei}) favours the assumption of a constant value for \leff. We plan to upgrade the LAr cell with PMTs of higher quantum efficiency and to improve significantly on the argon cleaning system.

LAr features the possibility of a separate study of singlet and triplet contributions in the domain of the luminescence quenching. Figure\,\ref{fig:leff} (right) shows a plot of the relative values \A/(\A+\B) and \B/(\A+\B) for nuclear and electronic recoils. The two energy scales determined in this work are displayed on top and bottom. Dashed lines correspond to linear fits to the individual values of \A\ and \B\ and are meant to guide the eye. In accordance to the conclusions of section\,\ref{cr}, changes in the relative strength of the components seem not to be located at low energies (where quenching effect become important) but are obvious for the whole region of study. Again, the measurements are compatible with an interpretation of collisional spin change of triplet state excimers or a preferred production of singlet states excimers during the self-trapping process of the excitons under high ionisation densities.

\vspace{-3mm}
\section*{Summary and conclusions}

We confirmed scintillation light quenching in liquid argon by impurities being of the same physical nature as observed in gaseous argon\,\cite{lumquench}. Similarly we use measured lifetimes and component ratios to reconstruct purity independent results. Further on we present preli\-minary results from data taken in LAr under excitation with mono-energetic neutrons, \alp-particles and photons at zero electric field. We confirm values for \CR\ monotonically increasing with values for \LET. Moreover changes of \CR\ seem to be decoupled from quenching processes and happen in regions with constant light yield per unit energy deposit. A component ratio close to the one expected from statistically populated singlet and triplet states (0.25) is observed for recoiling electrons in the region of lowest ionisation densities. For nuclear recoils we find \CR\ rising from values around 50\% to 75\% between 20 and 200\,keV. For \alp-particles in the MeV range we determine a relative scintillation yield \leffa\,=\,0.74$\pm$0.04 and observe a slightly smaller \CR\ as for recoiling nuclei. A preliminary analysis of the relative scintillation yield \leff\ for nuclear recoils at energies between 16 and 120\,keV shows a flat response within present errors. A mean value $<$\leff$>$\,=\,0.29$\pm$0.03 is found. No conclusive results for energies below that region can be drawn at the present state of the analysis. In the near future we plan to upgrade the cell with PMTs of larger quantum efficiencies and to improve the cleaning system. At a later stage we plan to add an internal electric field and to extract the ionisation charge (dual phase) to determine field and energy dependences in liquid argon of light and charge yields at working points relevant to dark matter searches. 


\vspace{-3mm}
\ack

This work is supported by grants from the Swiss National Science Foundation and the Aspera funded DARWIN project. 

\vspace{-4mm}
\section*{References}
\vspace{0mm}

\end{document}